# Extraction of Linear Carbon Chains Unravels the Role of the Carbon Nanotube Host

Lei Shi,[†] Kazuhiro Yanagi,[‡] Kecheng Cao,[¶] Ute Kaiser,[¶] Paola Ayala,[†] and Thomas Pichler*,[†]

†University of Vienna, Faculty of Physics, 1090 Wien, Austria

‡Tokyo ftetropolitan University, Department of Physics, 1-1 ftinami-Osawa, Hachiouji, Tokyo 192-0397, Japan

¶Ulm University, Central Facility for Electron fticroscopy, Electron fticroscopy Group of ftaterials Science, Ulm 89081, Germany

E-mail: thomas.pichler@univie.ac.at

## Abstract

Linear carbon chains (LCCs) have been shown to grow inside double-walled carbon nanotubes (DWCNTs) but isolating them from this hosting material represents one of the most challenging tasks towards applications. Herein we report the extraction and separation of LCCs inside single-wall carbon nanotubes (LCCs@SWCNTs) extracted from a double walled host LCCs@DWCNTs by applying a combined tip-ultrasonic and density gradient ultracentrifugation (DGU) process. High-resolution transmission electron microscopy (HRTEM), optical absorption, and Raman spectroscopy show that not only short LCCs but clearly long LCCs (LLCCs) can be extracted and separated from the host. Moreover, the LLCCs can even be condensed by DGU. The Raman spectral frequency of LCCs remains almost unchanged regardless of the presence of the outer tube of the DWCNTs. This suggests that the major importance of the outer tubes is making the whole synthesis viable. We have also been able to observe the interaction between the LCCs and the inner tubes of DWCNTs, playing a major role in modifying the optical properties of LCCs. Our extraction method suggests the possibility towards the complete isolation of LCCs from CNTs.

## Keywords

Linear carbon chains (LCCs) draw increasing attention because of their properties such as superior stiffness among the known materials and a direct electronic band gap,[1] which can be simply tuned by their length (from 4.0 to 1.8 eV), ranging from insulaing to semi-conducting.[2–5] Compared to other carbon allotropes much less could be understood in the years of research on LCCs, since these are extremely unstable and reactive. Therefore, LCCs are very difficult to synthesize,[6] and the longer ones with increased challenges. The high reactivity of LCCs represents an explanation to why they do not exist freely and should either be end-caped by chemical groups (big blocks)[7–11] or confined inside carbon nanotubes (CNTs)[12–17] to isolate them from each other and prevent their cross-linking. Although, these ending chemical groups or the hosting CNTs are indispensable for the integrality of LCCs, they affect their properties, chemical environment and bondings,[18–20] as well as confinement effects.[5,21] It is therefore important to understand this influence in detail. Inspired by the idea that longer chains tend to a more semiconductor behavior (the longer the LCC is, the lower the electronic band gap results),[5,21] historically the major goal across the field has been the production of the longest possible LCCs, but it took more than a century to achieve the polyyne consisting of from 2 to 44 carbon atoms by the end-caping method.[9] Further progress seemed always very difficult, and for that reason a new route has been highly desired for synthesis of longer polyyne.

CNTs have proven to be excellent nanoreactors inside which hollow core, structures like atomic wires[17,22–25] and graphene nanoribbons[26–28] can effectively grown. We recently demonstrated the successful synthesis of extremely long LCCs (LLLCCs) inside double-walled CNTs (DWCNTs) containing more than 6000 carbon atoms.[17] In this work we show that it is possible to extract the LLLCCs together with the inner tubes from the hosting outer tube of DWCNTs. This is a bridging step towards the complete extraction of LCCs from CNTs. The environmental effects on the LCCs are mainly ruled by the inner tubes of DWCNTs rather than the outer ones, making the modification of optical properties one of the most interesting outcomes of this extraction.

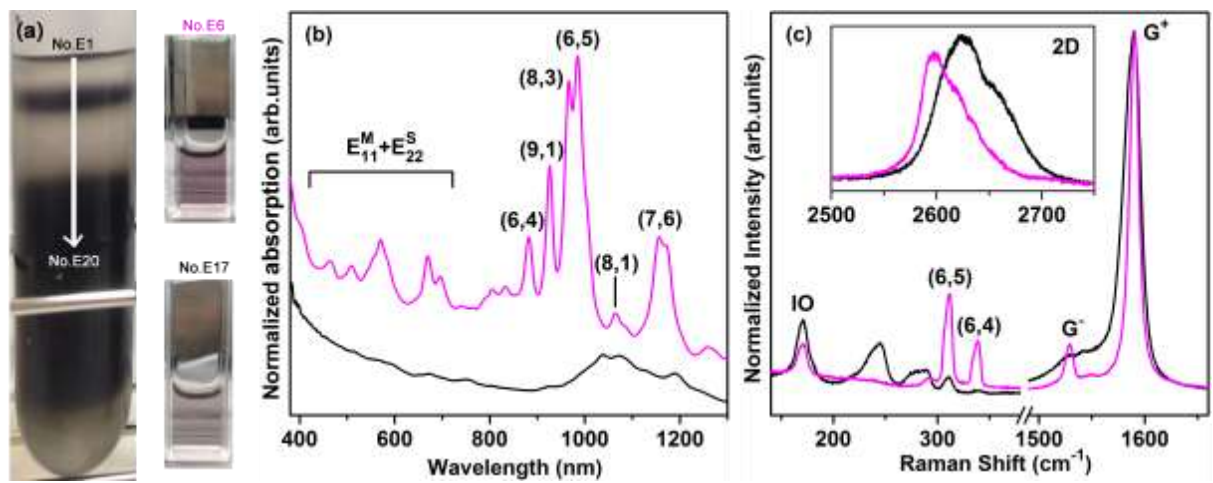

Figure 1: Separation of LCCs@SWCNTs and LCCs@DWCNTs by DGU. (a) Optical graph of a tube after DGU. The samples (0.7 mL for each) were collected from the top to the bottom marked by the extraction number from No.E1 to No.E20. The two photos show the No.E6 and No.E17 samples, which are in different color. Absorption (b) and Raman spectra (c) of the No.E6 and No.E17 samples both illustrate the successful extraction and separation. Inset of (c): 2D-bands.

## Results and discussion

The LCCs@inner tubes (LCCs@INCNTs) were extracted by tip-ultrasonication, and then DGU was applied to separate the extracted LCCs@INCNTs from the empty outer tubes as well as other remaining DWCNTs. As seen in Figure 1a, after the DGU process the LCCs@CNTs solution was clearly dispersed into two regions: The top thin layer (light purple) and the bottom part (light grey after dilution). The top region consists of thin SWCNTs as it is confirmed by the strong absorption spectrum plotted in Fig.1b. On the contrary, the absorption of bottom region is rather weak, similar as the reported for DWCNTs.[29,30] Note that we cannot recognize the contributions of the LCCs from the SWCNTs in this spectrum, since the weak absorption of LCCs overlapps and completely covered by the strong absorption of the SWCNTs.[31,32] The chirality assignment from the absorption shows that most of extracted tubes in the top layer are very small SWCNTs with diameters between 0.6 and 0.8 nm,[33] which are originally extracted from the inner tubes of DWCNTs. Furthermore, the Raman spectra shown in Fig.1c illustrate the difference between the two separated parts and confirm the successful extraction again. From these spectra (6,4) and (6,5) tube chiralities can be clearly identified in the low-frequency radial breathing mode (RBM) region for the top layer,[33–35] whereas much more signal from both the inner and outer tubes of DWCNTs can be seen in the bottom region sample. Note that the peak labelled IO corresponding to the signal arising from the iodixanol density gradient medium (see supplementary Fig.S1). The strong RBMs of (6,4) and (6,5) tubes is not only due to their enrichment in the sample (also see the absorption), but also because the tubes were excited by 568 nm laser whose energy is close to the nanotubes' $E_{22}$ resonance condition. [33,34] In addition, the narrower $G^+$-band, more pronounced $G^-$-band, and the shifted 2D-band (inset of Fig. 1c) indicate once again that only the small inner tubes exist in the top layer sample.[36]

In order to directly observe the extracted LCCs@INCNTs, No.E5 sample in the top region was examined by aberration-corrected transmission electron microscopy (ACTEM). Generally, the INCNTs are very long and usually formed into large bundles as shown in Fig. S2. If examined closely, the INCNTs indeed are SWCNTs, and the diameter of most of the INCNTs is between 0.6 and 0.7 nm. Two typical ACTEM images of parallel SWCNTs are shown in Fig. 2. In Fig. 2a, a short LCC with a length of 3.7 nm (approximately containing 30 carbon atoms) can be found in the upper SWCNT. The intensity valleys in the line profile of the upper SWCNT correspond to the LCC as well as the upper and lower walls of the host SWCNT, indicating that the LCC is confined in the SWCNT with a diameter of 0.64 nm (Fig. 2b and 2c). Moreover, another LCC with a length of 4.0 nm (approximately containing 32 carbon atoms) can be found in the middle SWCNT (0.62 nm) in Fig. 2d and 2e. It is a unforeseen observation that the chains survive after tip-ultrasonication and that the ACTEM observation is feasible, because even a large number of the small SWCNTs have been greatly destroyed. For example, the lower SWCNT in Fig. 2a was broken during the measurements by the electron irradiation (time-series ACTEM images are shown in Fig.S3), a conspicuous crack was found on the lower SWCNT in Fig. 2f pointed out by a green arrow, and a SWCNT was even unzipped into graphene nanoribbon as shown in Fig. S4.

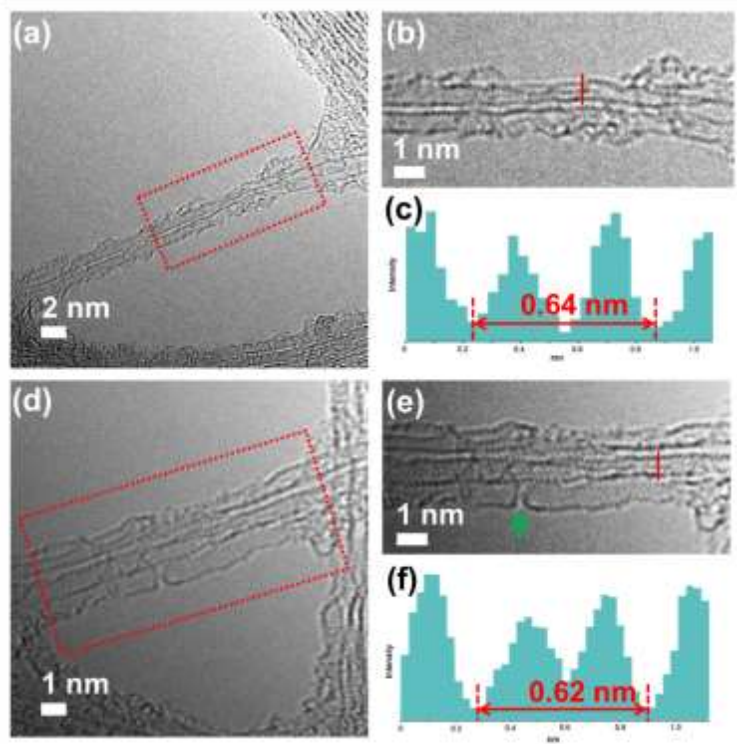

Figure 2: (a) A typical ACTEM image shows two parallel SWCNTs. A short LCC consisting of around 30 carbon atoms ( 3.7 nm) can be clearly observed inside the upper SWCNT. (b) Enlarged image of the wireframe area in (a). (c) Line profile at the position across the LCC@SWCNT marked by a red line shown in (b). (d) Another ACTEM image of three parallel SWCNTs. The LCC in the middle SWCNT consists of around 32 carbon atoms (4.0 nm). (e) Enlarged image of the wireframe area in (d). A crack pointed out by a green arrow exists in the lower SWCNT. (f) Line profile at the position across the LCC@SWCNT marked by a red line shown in (e).

Overall, the structure of extracted INCNTs is not intact and defects are frequently observed, therefore we did not observe any chains consisting of more than 100 carbon atoms even with the protection of the host SWCNT. Since TEM is a very local probe, the samples collected from different layers of the DGU tubes were examined in larger scale by absorption and Raman (macro mode with the laser spot size of about 5 mm) spectroscopies.

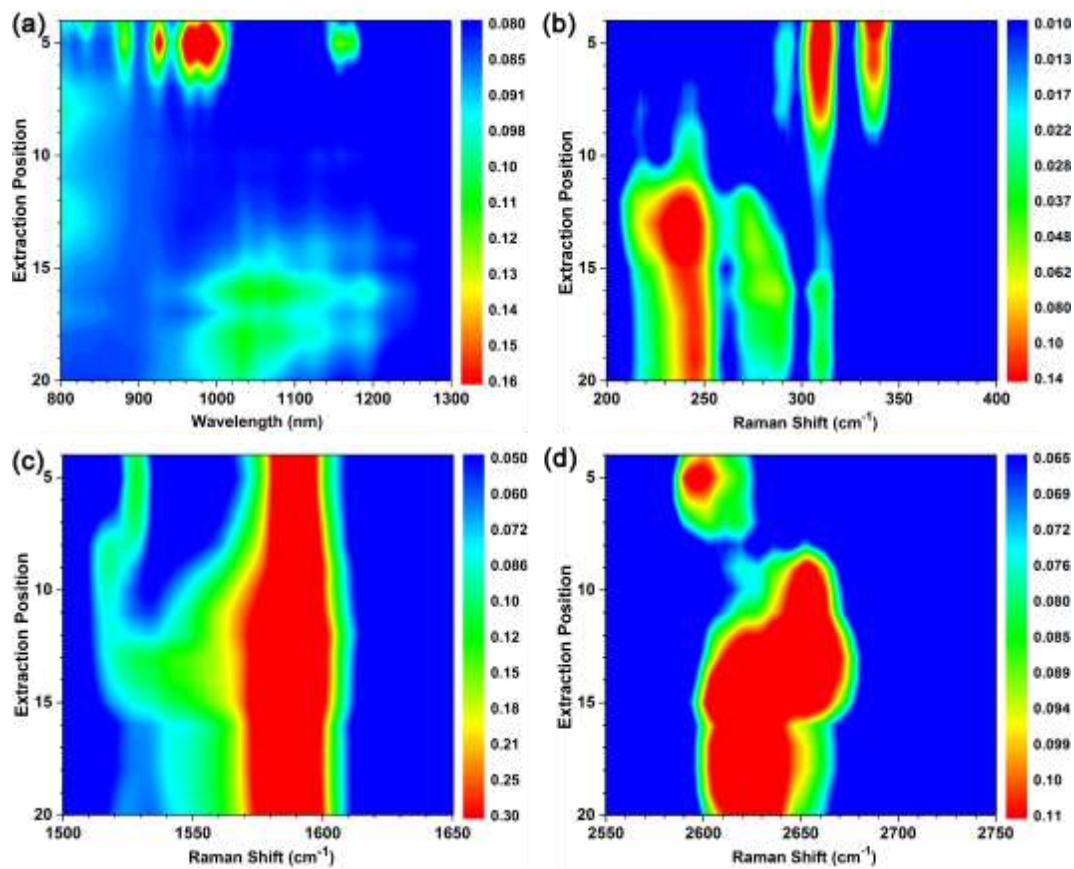

Figure 3: (a)Absorption contour mapping of the samples taken from No.4 to No.20 in the DGU tube. (b) RBM, (c) G-band, (d) 2D-band of Raman spectral contour mapping of the same layers. All of the spectral mapping confirm the successful extraction and separation of the LCC@SWCNTs from LCCs@DWCNTs.

The types of nanotubes as well as the LCC distribution can be both easily discerned by measuring the layers as marked by No.E1-No.E20 from the top to bottom in Fig.1a. The contour optical absorption in Fig.3a illustrates that the extracted inner tubes exist only in the top thin layer. More details about the types of nanotubes from the top to bottom can be obtained by Raman spectra as shown in Figure 3b-3d: Extracted inner tubes (No.4-8), outer tubes (No.E8-E11), mixture of outer tubes and DWCNTs (No.11-16), and DWCNTs (No.E16-E20) with overlapping on the boundaries. The outer tubes with intermediate density (the DWCNTs without extracted inner tubes) can be found in the layers No.E8-E15 by the following evidences: the much weaker intensity of RBM peaks between 300 and 350 $cm^{-1}$ corresponding to the inner tubes (Fig.3b); a broader G-band (Fig.3c); and the high ratio of the high-frequency component related to the large tubes in the 2D-band. From these, abundant outer tubes exist in the layers No.E8-E10. Importantly, the distribution of (6,5) in layers No.E4-E7 is almost constant (but with different concentration), whereas (6,4) gradually reduces from No.E4 to No.E7. As a result the distribution of different extracted inner tubes in the layers determines the LCC length distribution, as discussed below.

The DWCNTs and LCCs@DWCNTs were processed by the tip-ultrasonication with exactly the same parameters and then performing the DGU simultaneously. As shown in Fig.4, the Raman spectra of the extracted samples from the top region are exactly the same, except that the sample from LCCs@DWCNTs corresponds to LCCs inside the tubes (LCCs@SWCNTs). The contour optical absorption and Raman spectra of extracted and separated DWCNTs can be found in the supplementary Figs.S4-5. Furthermore, the overall extraction and separation are easily reproducible (see supplementary Figs.S6-7). The LCCs@SWCNTs were measured by two different lasers in order to excite LCCs with different lengths.[5] As shown in the lower insets of Fig.4, the LCC-band consists of a few components, corresponding to different length of LCCs. It is well understood that the longer the LCC, the lower the frequency, and the smaller electronic band gap.[4,5,37] Therefore, the 568 nm and 584 nm lasers are better for the detection of short and long chains, respectively. In addition, from the upper insets of Fig.4, the resonance behaviour has also been observed for (6,5) and (6,4) tubes.

The length distribution of LCCs was carefully checked by contour Raman spectroscopy. Two separated regions in Fig.5a are well illustrated for the DGU sample: LCCs@SWCNTs

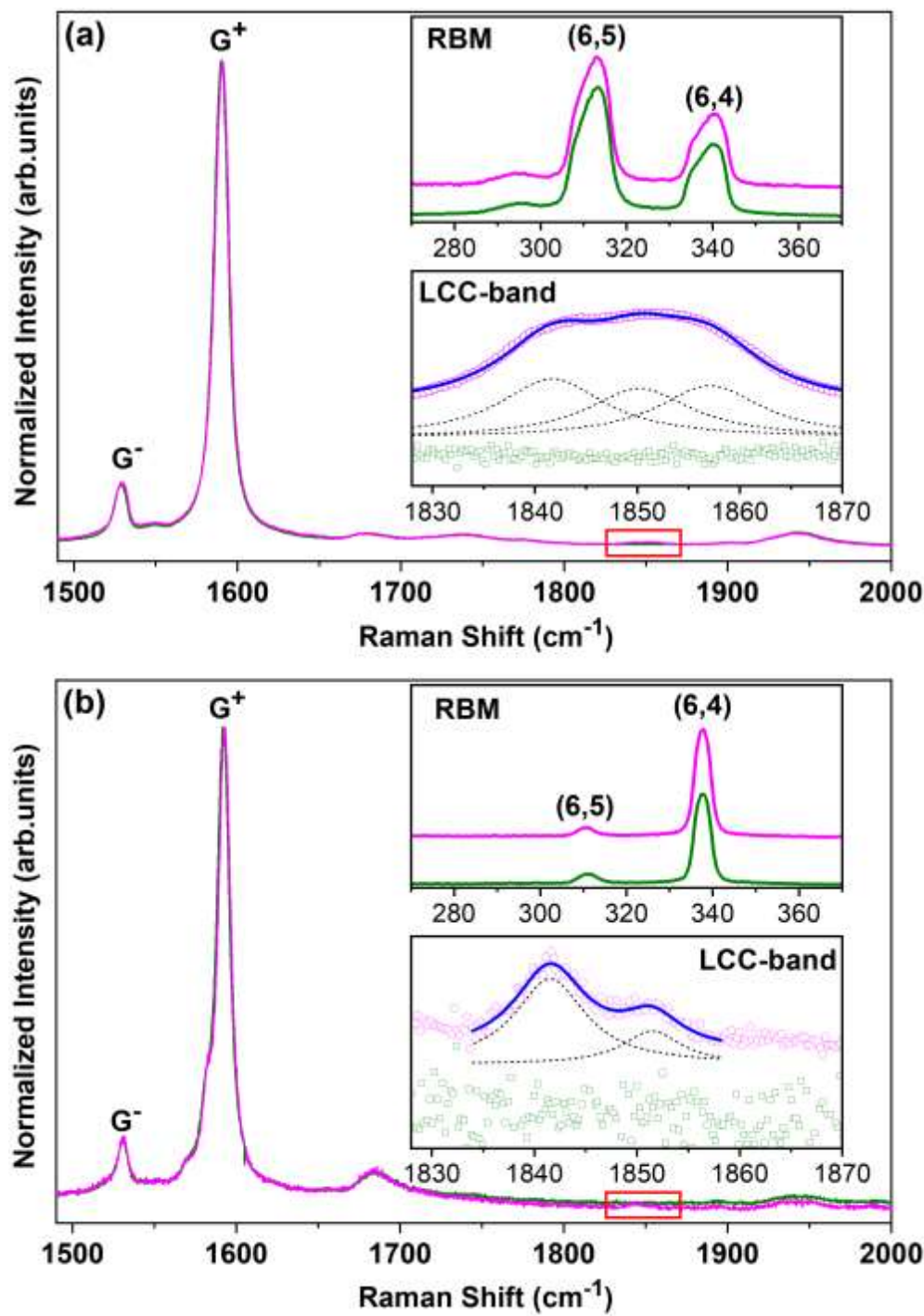

Figure 4: For comparison, pristine DWCNTs and LCCs@DWCNTs were both used for extraction and separation. Raman spectra of the extracted samples from the top layers of the DGU tubes show features of SWCNTs (Olive) and LCCs@SWCNTs (Magenta), excited by (a) 568 nm and (b) 584 nm lasers. Insets of (a) and (b): RBM (upper) and LCC-band (lower) of the extracted top layer samples. LCCs with different lengths between 1800 and 1870 $cm^{-1}$ were found in the extracted inner tubes.

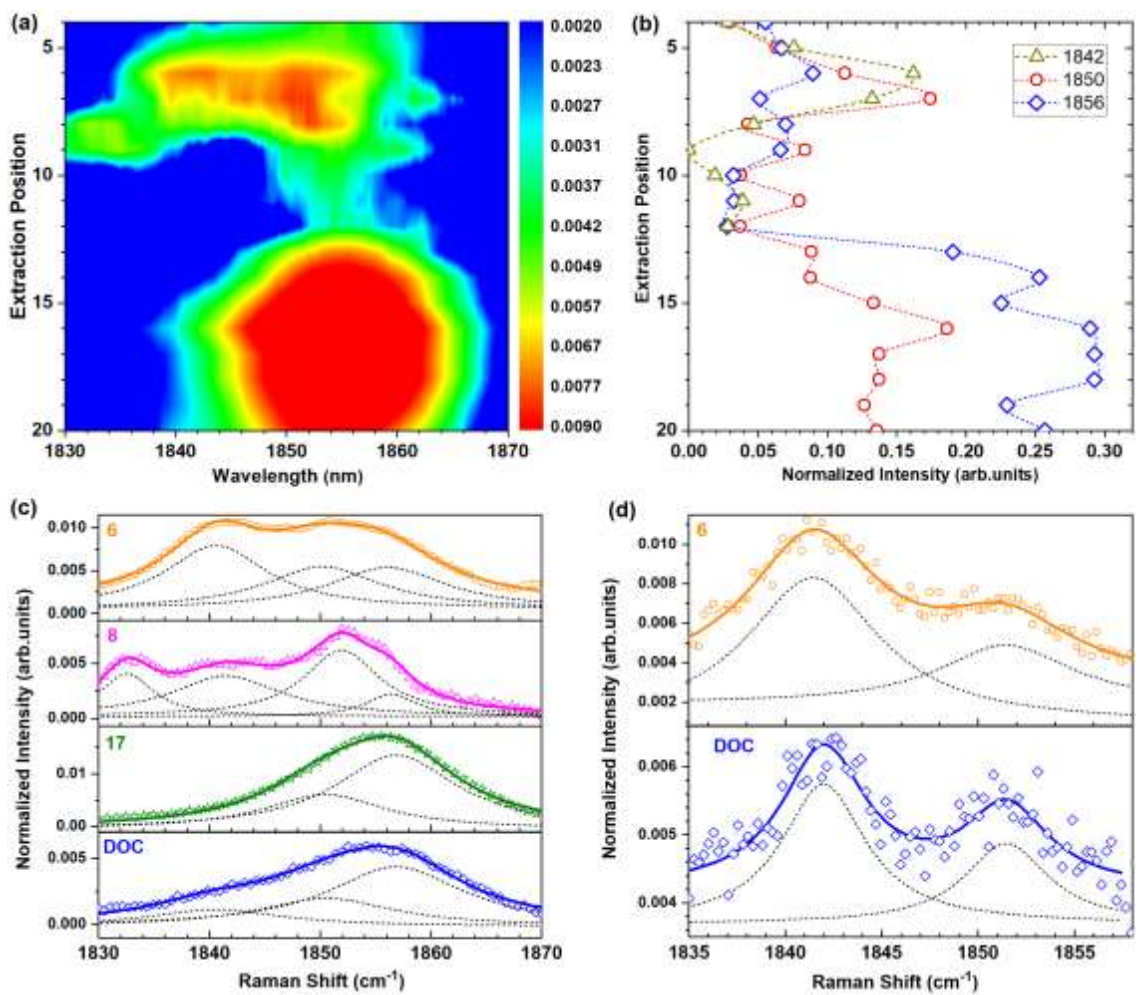

Figure 5: (a) LCC-band contour mapping of the samples from No.4 to No.20. (b) Raman spectral intensity of different peaks in the LCC-band obtained by peak fitting vs. number of the extracted samples. (c) LCC-band fittings of the Raman spectra excited by (c) 568 nm and (d) 584 nm lasers for some typical samples: No. 6, 8, 17, and the sample just after tip-ultrasonication in DOC solution (marked by DOC).

for the top and LCCs@DWCNTs for the bottom. Interestingly, the top region shows a larger contribution in the low-frequency area, revealing more longer LCCs in SWCNTs. If comparing the Raman spectrum of the sample just after tip-sonication (marked by DOC in Fig. 5c), we found the DGU separated samples No.E4-E20 all exhibit a higher intensity of LCC-band, which means the LCCs were concentrated by the DGU as well. This is understandable, because the LCCs can only be synthesized inside thin DWCNTs with an inner tube diameter of 0.6-0.85 nm, which corresponds to about 25 % in the DOC sample.[17]

The thin SWCNTs with LCCs have lower density hence stay at the higher region in the DGU tube, whereas high-density thick tubes without LCCs stay on the very bottom. In the top part of the DGU tube (Figs. 5b and 5c), long chains (1832 cm$^{-1}$) can be found in sample No.E8, more intermediate LCCs (1842 and 1850 cm$^{-1}$) are largely distributed in the samples No.E6 and E7, and only a few short chains (1856 cm$^{-1}$) are in the top-part samples. Detailed analysis of the ratio among long/intermediate/short LCCs in the separated samples is shown in Fig.5b. We found a few long LCCs and more intermediate LCCs are in the top-part sample, whereas more short LCCs are in the bottom-part samples. From the know-how developed on the LCC growth inside DWCNTs, it has been proved that thin DWCNTs are required for this growth to be feasible. However, this thin diameter must confine and protect the chains and protect the chains but it should not reach small diameter beyond a threshold at which no filling is feasible anymore. [17] This is consequent with the longer LCCs@thin SWCNTs (originally from long LCCs@thin DWCNTs) that appear in the top region after DGU, and vice versa. It is therefore helpful corroborating the ratio of intermediate and long LCCs present among the samples if excited by the 584 nm laser where they have a higher resonance. For example, about two times more intermediate LCCs can be found in the sample No.E6 than in the sample just after ultrasonication but before DGU (as marked by DOC in Fig.5d). A second DGU could be done subsequently to perform a fine separation on the LCCs@SWCNTs with different lengths, given that enough material can be obtained in the first DGU process.

Compared to the pristine solid LCCs@DWCNTs, [17] the LCC-band of the sample straight after tip-ultrasonication is very weak (Figs.1, 4, 5, and Figs.S7 and S8), which means that most of the LCCs have been destroyed by the tip-ultrasonication and its high power density. Some of the extracted inner tubes are also very defective, as seen in supplementary Fig.S3. Therefore, decreasing the power density could allow reducing the damage of the LCCs. Lower tip-ultrasonication power was applied with the same duration followed by the same DGU process. Indeed, the intensity of the LCC-band only decreased by about 5 times compared to the pristine LCCs@DWCNTs sample (Fig.6). The two separated regions in the DGU tube can be recognized as the LCCs@SWCNTs and the LCCs@DWCNTs. By comparison, the LCC-band intensity of the low-power extracted sample No.E3 is about 50 times higher than the intensity of high-power extracted sample (Fig.6c). Therefore, the extraction is successful as well at low power density with less damage, revealing that the extraction does not need much energy, given the superlubricity between the inner and outer tubes of DWCNT.[38–40] However, the low concentration of the extracted LCCs@SWCNTs by low power suggests longer tip-ultrasonication time could induce more optimal results.

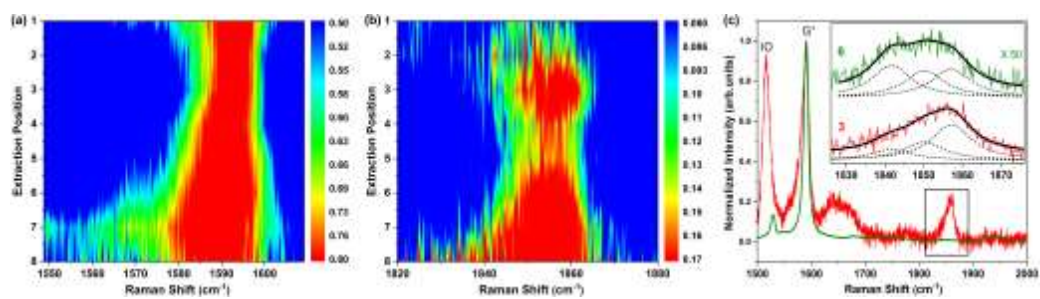

Figure 6: Lower tip-sonication power for extraction. Raman mapping of (a) G-band and (b) LLCC-band of the samples taken from the top to bottom in the DGU tube. (c) Raman spectra of two extracted samples both taken from the upper layers in the DGU tubes when using different tip-sonication power. Inset of (c): LCC-bands with fitting of the samples: No.6 from high power extraction and No.3 from low power extraction.

The LCCs@SWCNTs sample appears as stable as the LCC@DWCNTs, since we observe no changes on the Raman spectra and optical absorption from both LCCs@SWCNTs and LCCs@DWCNTs after storage under the ambient conditions for 2 years. The question then arise: what is the role of outer tube of DWCNT? Does it affect the LCCs? From the synthesis point of view, the outer tube is necessary to protect the thin inner tubes as nanoreactors for the growth of the LCCs, because the small SWCNTs with the same diameter as the inner tubes would be completely damaged at such high production temperature (around 1500 °C). However, as soon as the growth of LCCs finishes and they are ready for use, the outer tube is not indispensable like in the synthesis. It is has been reported that the interaction between the outer and inner tubes could shift the RBM of inner tubes by 30 cm$^{-1}$ at the maximum. [41,42] In our case, the frequency-shifts of the extracted (6,4) and (6,5) chirality tubes compared to those inside DWCNTs are both within 1 cm$^{-1}$ (Fig.1c and Figs.5c and 5d), suggesting that the effect of the outer tubes is comparable to the one from the surrounding IO/surfactants as well as the LCC in total. This effect resembles that of bundling regarding energy space increments and broadening of the van Hove singularities of the bundled tubes, explaining the shifts in the RBM frequencies. [43] Furthermore, if comparing the Raman spectra of extracted SWCNTs and LCCs@SWCNTs in Fig.4, almost no frequency shifts were observed for both (6,4) and (6,5) tubes. Previously, filled water molecules resulted in a RBM shift of the host nanotube by a few wavenumbers. [44,45] However, in our case the LCCs were only partly filled inside SWCNTs and have less than 1 % of carbon atoms in the whole system, thus they cannot affect the RBM mode of their host nanotubes. On the other hand, the LCCs could be certainly affected by the host tubes. We have shown that the Raman frequency of LCCs was red-shifted by their host tubes. [21] Therefore, we conclude that the outer tubes can not influence the LCCs, but the inner tubes do. Similarly, the screening effect on the carbon chains has also been attributed to the inner tube of the DWCNTs. [46]

## Conclusion

We have demonstrated the feasibility to extract LCCs@SWCNTs from LCCs@DWCNTs: Both short and long LCCs can be extracted, confirmed by powerful contour optical absorption and Raman spectroscopy. The DGU not only can separate the extracted LCCs@SWCNTs and the LCCs@DWCNTs, but also can concentrate the LCCs@SWCNTs. Although the strong tip-ultrasonication destroys most of the LCCs, decreasing the power of ultrasonication can decrease the damage ratio by about 50 times. Also, we confirmed that the interaction between the LCCs and the inner tubes affects the optical properties of the LCCs, whereas the outer tube does not. This interaction should be considered and overcome when extracting the LCCs completely out of the SWCNTs.

# Materials and Methods

## Extraction and Separation

The DWCNTs were synthesized by high-vacuum chemical vapor deposition and used as templates for synthesis of LLCCs by high temperature annealing at high vacuum.[17] The LLCCs@inner-tubes were extracted by tip-ultrasonication[31,47] and then separated from LL-CCs@DWCNTs by the DGU method, which previously was used to separate semiconducting and metallic tubes,[48] empty and water-filled SWCNTs,[49] as well as SWCNTs and DWC-NTs.[30] In short, 30 mg LLCCs@DWCNTs was ultrasonicated by bath-sonication in 30 mL toluene for 30 min, washed by methanol in a filtering system, tip-sonicated in 2 wt.% deoxy-cholate sodium salt (DOC, Tokyo Chemical Industry Co.) solution for 5 hours at 20 or 12 % output (Branson, 250DA), and centrifugalized at 36000 rpm. Top 2/3 of the supernatant were collected and used for further separation. The density gradient medium was prepared in a tube (40 PA seal tube, 345321A, Hitachi Koki Co.) by adding different concentrations of Iodixanol solution (IO in $H_2O$, OptiPrep$^{(TM)}$)) in a 2 wt.% sodium dodecyl sulfate (SDS, Sigma-Aldrich) solution from bottom to the top: 40 %, 35 %, 32.5 %, 30 %, 27 % of IO content (each for 7 mL), and finally the supernatant obtained in last step (6 mL). The tube was sealed and centrifuged at 50,000 rpm for 9 hours (Rotor P50VT2, Hitachi CP80WX). Finally, the separated solution was collected from the top to bottom layer by layer (each for 0.7 mL).

## Raman and Absorption Spectroscopic Measurements

The separated solution was measured under ambient conditions using a triple monochromator Raman spectrometer (Dilor XY with a liquid-nitrogen cooled Si CCD) excited by 568 and 584 nm lasers (R6G dye laser). The measurements were taken in macro-mode with 5 mm of laser spot and 4 mW of power. The spectral resolution is about 2 cm$^{-1}$. If not otherwise specified, all Raman spectra were normalized to the G$^+$-band.

The optical absorption was preformed by using a UV-VIS-NIR spectrophotometer (Shimadzu UV-3600). For ease of comparison, the spectra were normalized to their optical density at 900 nm.[32]

## Electron Microscopy Imaging

One drop of the No.5 solution consisting of LCCs@SWCNTs was put on the grid and rinsed throughly by hot water as well as ethanol on a filter with vacuum pump to remove the IO and other surfactants. To avoid excessive damage from the electron irradiation, aberration-corrected TEM (FEI Titan 80-300) with Cs-corrector was performed at 80 KV acceleration voltage with a modified filament extraction voltage for information limit enhancement (gun lens 3 and extention voltage 2000). Images were recorded on a slow-scan CCD camera type Gatan Ultrascsan XP 1000 (FEI Titan). Experimentally applied electron-fluxes ranged from $3 \times 10^6$ to $5 \times 10^6 e^-/nm^2/s$. TEM specimens were heated in air at 150 °C for 7 min shortly before insertion into the TEM column. All imaging experiments were carried out at room temperature.

## Acknowledgement

This work was supported by the Austrian Science Funds (FWF, P27769-N20) and the European Union's Horizon 2020 research and innovation programme under grant agreement No 664878. K.Y. acknowledges the JSPS KAKENHI through Grant Numbers JP16H00919, JP17H01069, JP17H06124. P.A. would like to acknowledge the contribution of the COST Action CA15107 (MultiComp).

# Supporting Information Available

Raman spectrum of IO/DOC/SDS solution, absorption and Raman mapping of extracted SWCNTs, absorption and Raman spectra of extracted LLCCs@SWCNTs from three different experiments, and Raman spectrum on solid LCCs@DWCNTs.

# Graphical  TOC Entry

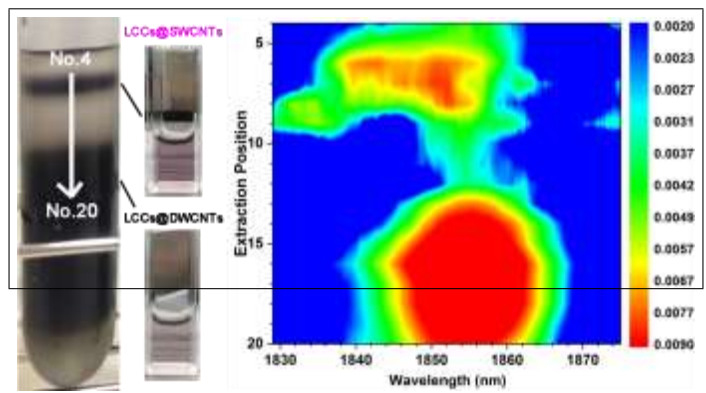